\documentclass[twocolumn,showpacs,preprintnumbers, amsmath,amssymb,amsfonts, final, a4paper,aps, citeautoscript,footinbib,prb,superscriptaddress]{revtex4-1}
\usepackage{graphicx}
\usepackage{tabularx} 
\usepackage{subfig} 
\usepackage{amsmath,amssymb,amsfonts,pifont,fancybox,float,mathtools}
\usepackage{graphicx}
\usepackage[vcentermath]{youngtab}
\usepackage{epsfig}
\newcolumntype{C}{>{\displaystyle}c<{}}
\newcolumntype{L}{>{\displaystyle}l<{}}
\newcolumntype{R}{>{\displaystyle}r<{}}
\usepackage[dvipsnames,usenames]{color}

\graphicspath{{fig/}}

\begin{document}
\title{Symmetry-protected topological phases in the SU($N$) Heisenberg spin chain: A Majorana-fermion approach}
\author{P. Fromholz}
\email{fromholz@ictp.it}
\affiliation{The Abdus Salam International Centre for Theoretical Physics (ICTP), Strada Costiera 11, 34151 Trieste, Italy\\}
\affiliation{Scuola Internazionale Superiore di Studi Avanzati (SISSA), via Bonomea 265, 34136 Trieste, Italy.\\}
\author{P. Lecheminant}
\affiliation{Laboratoire de Physique Th\'eorique et
Mod\'elisation, CNRS, CY Cergy Paris Universit\'e, 
95302 Cergy-Pontoise Cedex, France.\\}

\begin{abstract}
The nature of symmetry-protected topological phases of Heisenberg spin chains in totally symmetric representations of rank $N$ of the SU($N$) group is investigated through a Majorana fermion study starting from an integrable point.
The latter approach generalizes the one pioneered by Tsvelik [Phys. Rev. B {\bf 42},  10 499 (1990)] to describe the low-energy properties of the Haldane phase of the spin-1 Heisenberg chain from three massive Majorana fermions. We find, for all $N$'s, the emergence of a nondegenerate gapped phase with edge states whose topological protection depends on the parity of $N$. Whereas for $N$ odd, there is no such protection, the phase with even $N$ is shown to be topologically protected. We find that the phase belongs to the same topological class as the phase with edge states living in self-conjugate fully antisymmetric representation of the SU($N$) group.
\end{abstract}
\date\today
\maketitle     

\section{Introduction}

Majorana fermions, fermions that are their own antiparticles, have become one of the most important fundamental excitations of condensed-matter physics over the years.  A paradigmatic example is the one-dimensional (1D) Ising model in a transverse field which admits an exact description in terms of noninteracting Majorana degrees of freedom \cite{bookboso}.
These fermions have a nonlocal character in terms of the underlying spins since they can be viewed as the bound states of a local spin flip and a domain-wall topological defect. This Majorana approach gives a full description of the properties of the Ising quantum critical point that defines the simplest conformal field theory (CFT) with central charge $c=1/2$ \cite{bookboso,dms}.
These fermions experience only fermion number parity conservation, a ${\mathbb{Z}}_2$ symmetry. Yet, several copies of such degrees of freedom allow the investigation of more complicated situations with a continuous symmetry. One striking example is the study of the confinement of fractional quantum numbers that occur in weakly coupled two-leg spin-1/2 Heisenberg ladder. 
The low-energy excitations of this system can be mapped onto non-interacting four massive Majorana fermions \cite{bookboso,shura}.

The second interest in Majorana fermions lies at the heart of exotic physics. It stems from the formation of zero-energy Majorana modes that are localized around specific points with  topological features, such as domain walls, vortices, or boundaries.
The onset of non-Fermi liquid behavior in the two-channel Kondo problem where a spin-1/2 impurity spin is located on a metal with two degenerate channel degrees of freedom has been described within the Toulouse-limit solution of the model as due to the presence of a localized Majorana fermion \cite{emery}. The electronic channels overscreen the impurity spin, and a zero-energy Majorana mode located at the impurity is decoupled from the conduction degrees of freedom, giving rise to a finite ground-state entropy $\ln \sqrt 2$.  These Majorana zero modes, which are not particles and not even fermions, have intriguing quantum properties with non-Abelian anyon statistics, ground-state degeneracy, and robustness. In this respect, they have promising applications to topological quantum information processing \cite{dasarma}.

The simplest 1D model with Majorana zero modes is the Kitaev chain which is a 1D lattice version of a spinless $p$-wave superconductor \cite{kitaevmajorana}. The model has a topologically protected gapful phase that hosts an unpaired Majorana zero mode at the two ends of the chain. The Kitaev chain with time-reversal symmetry belongs to the BDI class of the tenfold classification of noninteracting topological insulators and superconductors \cite{10fold}.  
This BDI  class is characterized by a ${\mathbb{Z}}$-valued topological invariant. The ${\mathbb{Z}}$-valued topological invariant can be incremented by stacking an additional Kitaev chain to the system.  This noninteracting  ${\mathbb{Z}}$  classification of  BDI class is reduced to ${\mathbb{Z}}_8$ in the presence of interactions \cite{fidkowski,kitaev,turner}.
A topological phase with eight Majorana zero modes at the two ends of the chain is adiabatically connected by interactions to a gapful featureless phase without closing the bulk gap.  

Majorana fermions and Majorana zero modes may represent an avenue to describe 1D bosonic interacting symmetry-protected topological (SPT) phases. The latter denomination refers to 1D nondegenerate gapped phases of spins or bosons whose edges states are protected by a given symmetry. These phases with on-site protecting symmetry group G 
are known to be classified by the second cohomology group $H^{2}(G, \text{U(1)})$ which labels the inequivalent projective representations of the symmetry at the edge, i.e., the nature of the boundary spin \cite{kitaev,wenchen,cirac}.
The Haldane phase \cite{haldane} of the spin-1 Heisenberg chain is a paradigmatic example of  a 1D interacting SPT phase with its Haldane gap and the existence of  spin-1/2 edge states that can be simply understood from the Affleck-Kennedy-Lieb-Tasaki (AKLT) approach \cite{aklt}.  Here, in the presence of an internal rotation G= SO(3) symmetry, there is a 
$ H^2({\rm SO(3)}, U(1)) = {\mathbb{Z}}_2$ classification. The Haldane phase is, thus, the only SO(3) SPT phase with its edge states that transform projectively in the spinorial representation of the SO(3) group, i.e., the spin-1/2 representation of SU(2)\cite{oshikawapollmann}.

Several approaches captures the main physical properties of the Haldane phase. Indeed, the phase can be described by staking four copies of the Kitaev chain \cite{verresen,xu}. An O(4) symmetry emerges by construction from which an SO(3) subgroup acts projectively on the boundary. A second more conventional approach was pioneered by Tsvelik in Ref.~\onlinecite{tsvelik}.  This approach describes the Haldane phase starting from an integrable spin-1 model, the Babujian-Takhtajan (BT) model~\cite{babujian}, whose critical properties are governed by three decoupled gapless Majorana fermions. A deviation from this integrable point leads to the formation of a gap. For a semi-infinite chain, three Majorana zero modes emerge at the edge \cite{orignac}. These modes generate the spinorial representation of SO(3) and, thus, lead to the spin-1/2 edge states of the Haldane phase \cite{orignac}.  A third alternative approach is the well-known semiclassical description of the Haldane phase by the O(3) nonlinear $\sigma$ model with a $\theta = 2 \pi$ theta term \cite{haldane}.
The precise value of this topological angle leads to the liberation of spin-1/2 edge states when the chain is opened \cite{ng}.

In this paper, we investigate the possible Majorana fermion description of 1D SPT phases protected by a higher continuous symmetry group G.  A known example is when G $=$ SO($2n+1$). Then, $ H^2({\rm SO(2n+1)}, U(1)) = {\mathbb{Z}}_2$ revealing the SPT phase that generalizes the Haldane phase for $n>1$~\cite{tu2}. Its physical properties can be described by $2n+1$ massive noninteracting Majorana fermions by exploiting the existence of an integrable model with SO($2n+1$) symmetry \cite{nonneSO5,tu1}. In a semi-infinite geometry, $2n+1$ Majorana zero modes are located at the boundary and give a ground-state degeneracy of $2^{n}$ which is the dimension of the spinorial, i.e., projective, representation of the  SO($2n+1$) group  \cite{nonneSO5,tu1}. The main properties of the SO($2n+1$) SPT phase are then reproduced by means of this Majorana fermion  approach.

What happens if we consider richer 1D SPT phases when the on-site protection symmetry G is the projective unitary group $\text{PSU($N$)} \simeq \text{SU($N$)}/\mathbb{Z}_N $ ? Since $H^{2}(\text{PSU($N$)}, \text{U(1)}) = \mathbb{Z}_N$, 
$N-1$ interesting SPT phases are expected that are protected by the $\text{PSU($N$)}$ group or its discrete subgroup 
$\mathbb{Z}_N \times \mathbb{Z}_N$ \cite{quella, quella1, else}.  
Microscopic realizations of these phases appear in the SU($N$) antiferromagnetic Heisenberg spin chain, 
\begin{equation}
{\cal H} = J \sum_i \sum_{A=1}^{N^2 -1} S^{A}_i S^{A}_{i+1},
\label{Heisenberg}
\end{equation}
where the spin operators $S^{A}_i $ on each site $i$  of the chain belong to a given irreducible representation 
of the SU($N$) group which is described by a Young tableau with $n_Y$ boxes. 

There is now a rather good understanding of the physical properties of model (\ref{Heisenberg}). Some of them and 
related topics are reviewed in Ref.\onlinecite{capponireview}. The generalization of the Haldane conjecture for SU($N$)  
is described by three different cases depending on the value of $n_Y$ with respect to $N$ \cite{greiter,affleckmila1,affleckwamer}.
When $n_Y$ and $N$ are coprime, both a semiclassical approach of model (\ref{Heisenberg}) in Refs. \onlinecite{affleckmila1,affleckmila,affleckwamer} and a CFT analysis \cite{lecheminant2016,oshikawa} have shown 
that a quantum critical behavior in the SU($N$)$_1$ universality class with central charge $c=N-1$ emerges. 
In contrast, when $n_Y$ and $N$ have a nontrivial common divisor different from $N$, a spectral gap is formed \cite{greiter,oshikawa,affleckwamer}. The one-step translation symmetry T$_{a_0}$ of model  (\ref{Heisenberg}) is spontaneously broken resulting in a ground-state degeneracy.
The last case, the most interesting for us, is when $ n_Y = 0 \; \; {\rm mod}   \; N$ and a Haldane gap phase is expected \cite{greiter}. For these representations, the continuous symmetry group of model (\ref{Heisenberg}) is the projective unitary group $\text{PSU($N$)}$ and the $N-1$ different SPT phases might be found in the lattice model (\ref{Heisenberg}). Their edge states are labeled by the inequivalent projective representations of $\text{PSU($N$)}$, which are specified by $\mathbb{Z}_N$ quantum numbers $n_{\rm top} = n_{Y\rm edge}  \; {\rm mod}   \; N$, $n_{Y\rm edge}$ being the number of boxes of the Young tableau corresponding to the representation of the boundary spins. 

Several SPT phases have already been identified in the PSU($N$) Heisenberg chain (\ref{Heisenberg}).
The topological class with $N$ even and $n_{\rm top} = N/2$ appears when the spins on each site belong to 
the representation with the Young tableau \cite{Nonne2013,bois,totsuka},
\begin{equation}
\text{\scriptsize $N/2$} \left\{ 
\yng(2,2,2,2)
 \right.   \; .
 \label{sptnonne}
 \end{equation}
The edge state belongs to the self-conjugate fully antisymmetric representation of the SU($N$) group such that
$ n_{Y\rm edge} = N/2$. For $N=3$ and $N=4$, the remaining SPT phases are the chiral SPT phases $({\bf N}, {\bf \bar N}), ({\bf \bar N}, {\bf  N})$. For instance, $({\bf N}, {\bf \bar N}$) denotes a nondegenerate fully gapped phase such that the left (respectively, right) edge state transforms in the fundamental representation ${\bf N}$ (respectively, antifundamental ${\bf \bar N}$) of the SU($N$) group. These two chiral SPT phases are the two ground states of the model (\ref{Heisenberg}) in the adjoint representation \cite{AKLTlong,schuricht,furusaki,katsura,quella2}.
All these PSU(3) and PSU(4) SPT phases have been realized in lattice systems of ultracold fermions loaded into optical lattices or in spin-ladder systems  \cite{Nonne2013,bois,furusaki,fromholz2019,momoi,fromholz2020}.

In this paper, we consider the general PSU($N$) case by focusing on the symmetric rank-$N$ tensor representations, described by a Young tableau with $N$ boxes  and a single line: $ \yng(2).. \yng(1)$. Since $n_Y = N$, the emergent phase is a good candidate for being a SPT phase. When topological, the phase constitutes the natural generalization of the Haldane phase for $N > 2$ within the PSU($N$) series as $\text{SO(3)}  \simeq \text{SU($2$)}/\mathbb{Z}_2 \simeq  \text{PSU(2)} $. In the following, we develop an approach for generic $N$ to describe the possible SPT phase in terms of $N^2-1$ massive Majorana fermions and their associated zero-Majorana modes for a semi-infinite chain. An even-odd effect is found.
The ground state of model (\ref{Heisenberg}) in the symmetric rank-$N=2n$ representation is shown to describe a stable SPT phase with topological index $ n_{\rm top} = n$ which shares the same topological class as the SPT phase of the PSU($2n$) Heisenberg chain in the representation (\ref{sptnonne}). When $N$ is odd, there is no such protection and the phase can be adiabatically connected to a  trivial gapful featureless phase without closing the bulk gap.  

The remainder of this paper is organized as follows. In Sec. II, we present our low-energy approach to describe the properties of model (\ref{Heisenberg}) in the symmetric rank-$N$ representation starting from an integrable spin model. 
In Sec. III, we exploit a conformal embedding onto $N^2-1$ gapless Majorana fermions. This embedding leads to the emergence of a PSU($N$) SPT phase whose boundary spin is described in terms of $N^2-1$ zero Majorana modes.
Finally, Sec. IV summarizes our findings and the Appendix presents the AKLT construction of the model for $N=4$.

\section{Low-energy approach}

In this section, we present our strategy to develop a field-theory analysis for describing the fully gapped phase of 
the PSU($N$) Heisenberg spin chain (\ref{Heisenberg}) for symmetric rank-$N$ tensor representation.

\subsection{Integrable SU($N$) spin model}
The starting point of the analysis is the existence of an integrable SU($N$) model
with degrees of freedom in symmetric rank-$k$ tensor representation, introduced by Andrei
and Johannesson (AJ) \cite{andreiJ,johannesson}.
The AJ model involves a specific polynomial $P(x)$  of degree $k$ in terms of the bilinear term $S^{A}_{i}  S^{A}_{i+1}$,
\begin{equation}
{\cal H}_{\rm AJ} = J \sum_i \sum_{A=1}^{N^2 -1} P(S^{A}_i S^{A}_{i+1}).
\label{AJlattice}
\end{equation}
The explicit expression of the polynomial is not important for this paper and can be found in Ref. 
\onlinecite{johannesson}. Model (\ref{AJlattice}) is the SU($N$) generalization of Bethe-ansatz integrable spin-$S=k/2$ Heisenberg chain models which display a gapless behavior described by the SU(2)$_{2S}$ CFT  \cite{affleckschulz, affleckhaldane}. For $N=2$ and $k=2$, the AJ model reduces to the BT spin-1 model with Hamiltonian \cite{babujian},
\begin{equation}
{\cal H}_{\rm BT}=J\sum_i \left[ \mathbf{S}_i\cdot\mathbf{S}_{i+1} + \beta
(\mathbf{S}_i\cdot\mathbf{S}_{i+1})^2\right],
\label{biquaham}
\end{equation}
with $\beta = -1$ and ${\bf S}_{i}$ is a spin-1 operator at site $i$. 

The main bulk properties of the Haldane phase of the spin-1 Heisenberg chain have been derived by Tsvelik in Ref. \onlinecite{tsvelik} by introducing a small deviation $\beta = -1 + \delta (0 < \delta \ll 1) $ from the SU(2)$_2$ critical point of the BT model (\ref{biquaham}). Starting from this critical point with central charge $c=3/2$, which can be described in terms of three massless Majorana fermions, it was shown that the low-energy properties of  the Heisenberg spin-1 chain could be captured by a triplet of noninteracting massive Majorana fermions.  Later, the hallmark of the Haldane phase, i.e., its spin-1/2 edge state, has been derived within this field theory analysis \cite{orignac}. 

Our aim, here, is to present the generalization of Tsvelik's approach to describe the possible formation of an SPT phase in the Heisenberg spin-chain model (\ref{Heisenberg}) for the specific symmetric rank-$N$ tensor representation starting from the integrable AJ spin model. It has been shown numerically that the AJ model displays a quantum critical behavior in the SU($N$)$_k$ universality class with central charge $c= k(N^2-1)/(N+k)$ \cite{alcaraz,rachel}.  In the special $k=N$ case,  $c=(N^2-1)/2$ which is the central charge of $N^2-1$ gapless Majorana fermions.
The low-energy properties of the AJ model for $k=N$ are  described by the SU($N$)$_N$ Wess-Zumino-Novikov-Witten (WZNW) model \cite{dms,bookboso} perturbed by a marginal irrelevant current-current interaction with Hamiltonian density \cite{affleck,lecheminantreview}, 
\begin{equation}
{\cal H}_{\rm AJ} =  \frac{\pi v}{N}\Big(:J_R^A J_R^A: + :J_L^A J^A_L:\Big) + \gamma \;  J_R^A J_L^A,  
\label{AJmodelcont}
\end{equation}
where $v$ is the spin velocity, $:O:$ denotes the normal ordering of operator $O$ and a summation over repeated
SU($N$) indices $A= 1, \ldots, N^2 -1$ is assumed in the following. In Eq. (\ref{AJmodelcont}), $J_{R,L}^A$ are the chiral currents which satisfy the SU($N$)$_N$ current algebra,
\begin{eqnarray}
J_{L}^A\left(z\right) J_{L}^B\left(\omega\right) &\sim& \frac{N \delta^{AB}  }{8 \pi^2 
\left(z - \omega \right)^2} + 
\frac{i f^{ABC} }{2 \pi \left(z - \omega \right)} J_{L}^C\left(\omega\right),
\label{curralgSUNk}
\end{eqnarray}
with a similar definition for the right current. In Eq. (\ref{curralgSUNk}), $f^{ABC}$ denotes the antisymmetric structure constants of the SU($N$) group and $z = v \tau + i x$ ($\tau$ being the imaginary time). The marginal irrelevant term with $\gamma<0$ of Eq. (\ref{AJmodelcont}) accounts for logarithmic corrections  of the  SU($N$)$_N$  quantum criticality \cite{affleck,itoi}. 
The critical point with $\gamma =0$ is described by the SU($N$)$_N$ WZNW model with Euclidean action  \cite{witten,knizhnik}, 
\begin{eqnarray}
&& {\cal S}_{\rm WZNW} = \frac{N}{8\pi} \int_{M_2} d^2 x \; {\rm Tr} \; (\partial_{\mu} G^{\dagger} \partial_{\mu} G) 
\nonumber \\
&+& \frac{i N }{12\pi}   \int_{M_3} d^3 y \; \epsilon^{\alpha \beta \gamma} 
 {\rm Tr} \; (G^{\dagger} \partial_{\alpha} G G^{\dagger} \partial_{\beta} G G^{\dagger} \partial_{\gamma} G),
\label{WZW}
\end{eqnarray}
$G$ being an SU($N$) matrix field, the WZNW field, and $M_3$ is a three-dimensional manifold whose boundary is the two-dimensional Euclidean space: $\partial M_3 = M_2$.  The critical point of the AJ model is expected to be  fragile on general grounds due to its fine-tuning \cite{affleck}.  A fully gapped PSU($N$) SPT phase might emerge in the close vicinity of the integrable AJ point in close parallel to the $N=k=2$ case. A perturbed SU($N$)$_N$ CFT would then explain the formation of the spectral gap and the low-energy properties. 

Our next task is to identify the suitable relevant perturbation which accounts for the emergence of a non-degenerate fully gapped phase for model (\ref{Heisenberg}) where the spin operators belong to symmetric rank-$N$ tensor representation of SU($N$). 

\subsection{Departure from the SU(N)$_N$ WZNW fixed point}

The allowed strongly relevant operators which control the departure from the SU($N$)$_N$ quantum critical point should be invariant under the symmetries of the underlying lattice model (\ref{Heisenberg}).  The most important lattice symmetry is the one-step translation invariance T$_{a_0}$ which corresponds to a ${\mathbb{Z}}_N$ symmetry in
the continuum limit. This correspondence stems from the underlying U($N$) fermionic Hubbard model of model (\ref{Heisenberg}) or  (\ref{AJlattice}) which is at $1/N$ filling with Fermi momentum $k_F = \pi/N a_0$, $a_0$ being the lattice spacing \cite{affleck,lecheminantreview}.
The WZNW primary field $G$ of Eq. (\ref{WZW}), transforming in the fundamental representation of SU($N$), takes the following form under T$_{a_0}$ \cite{affleck,lecheminantreview}:
\begin{equation}
G \rightarrow \omega G ,
\label{translation}
\end{equation}
with $\omega = e^{ 2 i \pi/N}$.
The spectrum of the SU($N$)$_N$  CFT is described by primary operators 
which transform in a limited set of representations of the SU($N$) group. The highest-weights ${\Lambda} = (\lambda_1 \lambda_2 \ldots  \lambda_{N-1})$ of these representations must satisfy the constraint: $ \sum_{i=1}^{N-1} \lambda_i   \le N$, $\lambda_i$ being the Dynkin labels. Introducing $l_i = \sum_{j=i}^{N-1}  \lambda_j$ as a Young tableau row length, we see that the Young tableau cannot have more than $N$ columns. The scaling dimensions of the
primary fields are related to the quadratic Casimir of the underlying representation of  SU($N$) 
\cite{knizhnik,zamoloWZW,schnitzer},
\begin{eqnarray}
\Delta_{\Lambda} = \frac{X + n_{Y} (N+1) - n_{Y}^2/N}{2N},
\label{scalingdimSUN}
\end{eqnarray}
with $n_{Y} =  \sum_{i=1}^{N-1} l_i = \sum_{i=1}^{N-1} i \lambda_i$ which is the number of boxes in the Young tableau 
and $X =   \sum_{i=1}^{N-1}   l_i ( l_i - 2 i)$.

The possible allowed primary fields, obtained from fusion of the fundamental field $G$, should be invariant under T$_{a_0}$. From Eq. (\ref{translation}), we observe that they transform under representations of SU($N$) which are described by a Young tableau such that $n_{Y}$ is a multiple of $N$. For instance, all primary fields which transform according to totally antisymmetric representations of SU($N$), $\Lambda^l {\bf N}$ ($l=1, \ldots, N-1$), cannot appear in the continuum description of the Heisenberg spin chain model (\ref{Heisenberg}) since they acquire a phase factor $e^{ 2 i l \pi/N}$
under T$_{a_0}$. The most relevant operator, which is translational invariant, turns out to be the primary field in the adjoint representation with highest weight: (1 0 \ldots 0 1). The latter can be expressed in terms of the SU($N$)$_N$ WZNW field $G$\cite{knizhnik},
\begin{equation}
\Phi^{AB}_{\rm adj} \sim {\rm Tr} ( G^{\dagger} T^{A} G T^{B} ), 
\label{adjointfield}
\end{equation}
$T^{A}$ being the SU($N$) generators transforming in ${\bf N}$, normalized such that $ {\rm Tr} (T^{A} T^{B}) =  \delta^{AB}/2$. According to Eq. (\ref{scalingdimSUN}), the scaling dimension of the adjoint primary SU($N$)$_N$ field is 
$\Delta_{\rm adj} = 1$ for all $N \ge 2$. For $N=3$, the subleading translation-invariant primary fields are marginal ($\Delta_{\Lambda}=2$).  They transform in the ${\bf 10}$ and  $\mathbf{\overline{10}}$ representations with the Young tableaux,
\begin{align*}
{\bf 10} & \longleftrightarrow  \yng(3) \;,\\
\mathbf{\overline{10}} & \longleftrightarrow  \yng(3,3) \;.
\end{align*}
In the SU(4) case, there is a subleading relevant primary field with scaling dimension $3/2$ which transforms 
in the self-conjugate ${\bf 20}$ representation of SU(4) with:
\begin{equation*}
{\bf 20} \longleftrightarrow \yng(2,2) \; .
\end{equation*}
The remaining allowed operators are marginal or irrelevant. Our numerical analysis for $N \ge 4$ shows that only the adjoint operator and one other primary field $\Phi^{'}$ are both strongly relevant and translation invariant. The latter transforms in the self-conjugate representation of  SU($N$) with the Young tableau of $N$ boxes,
\begin{equation}
\text{\scriptsize $N-2$} \left\{ 
\yng(2,2,1,1)
 \right.   \; .
 \end{equation}
The primary field $\Phi^{'}$ has scaling dimension $\Delta^{'} = 2 (N-1)/N < 2$ and is, indeed,
a relevant contribution but less relevant than the adjoint field. Such an operator is, in fact, generated by the fusion of the adjoint field by itself,
\begin{equation}
\Phi_{\rm adj} \times \Phi_{\rm adj} \sim I + \Phi_{\rm adj} + \Phi^{'}+..,
\label{fusionrule}
\end{equation}
where the dots describe terms that are marginal or irrelevant operators. 

The leading Hamiltonian density  which describes the departure from the AJ model reads, thus, as follows  depending on $N$,
\begin{subequations} \label{modelcont}
\begin{align}
\begin{split}
  {\cal H}_{N=2,3} =&  \frac{\pi v}{N}\Big(:J_R^A J_R^A: + :J_L^A J^A_L:\Big) \\
  & \quad + g \; {\rm Tr}  \Phi_{\rm adj},  
\end{split}\\
\begin{split}
  {\cal H}_{N \ge 4} =&  \frac{\pi v}{N}\Big(:J_R^A J_R^A: + :J_L^A J^A_L:\Big) \\
& \quad + g \; {\rm Tr}  \Phi_{\rm adj} + \lambda \; {\rm Tr}  \Phi^{'} ,\label{modelcontb}
  \end{split}
\end{align} 
\end{subequations}
where we have neglected marginal and irrelevant perturbations. The SU($N$)$_k$ CFT perturbed by the adjoint primary field has been investigated  and a massless flow to SU($N$)$_1$ is expected when $N$ and $k$ have no common divisor \cite{lecheminant2016}. In Eqs. (\ref{modelcont}), we have $k=N$, and it is then likely as it will be shown below that a spectral gap is formed. We, thus, expect that the physical properties of the PSU($N$) Heisenberg spin chain (\ref{Heisenberg}) in symmetric rank-$N$ tensor representation of SU($N$) are captured by the low-energy theory (\ref{modelcont}). 

\subsection{WZWN model and $\sigma$ model on a flag manifold}

We now switch to a Lagrangian description to study the infrared properties of the perturbed CFT (\ref{modelcont}) and to make a connection to the semiclassical field theory derived recently for SU($N$) Heisenberg chain (\ref{Heisenberg}) in totally symmetric representations \cite{bykov,bykov2,affleckmila,affleckmila1}.
In this respect, we consider the following action first introduced in Ref. \onlinecite{seiberg}: 
\begin{eqnarray}
{\cal S} =  {\cal S}_{\rm WZNW} + \sum_{n=1}^{\left[N/2\right]} \int_{M_2} d^2 x  \; g_n {\rm Tr} \left[ G^n\right] {\rm Tr}  \left[\left(G^{\dagger}\right)^n\right],
\label{WZWint}
\end{eqnarray}
where $n=1$ and $n=2$ potential terms correspond to the two relevant operators of Eq. (\ref{modelcont}) for $N > 3$ since 
Eq. (\ref{adjointfield}) imposes 
\begin{equation}
{\rm Tr}  \Phi_{\rm adj} =  {\rm Tr} G  {\rm Tr} G^{\dagger}-  {\rm Tr} (G^{\dagger} G)/N \sim  {\rm Tr} G  {\rm Tr} G^{\dagger},
\label{tradj}
\end{equation}
$G$ being an SU($N$) matrix in  the Lagrangian approach, whereas $\Phi^{'}$ appears in the fusion $G^2 \otimes (G^{\dagger})^2$. 

Let us first discuss the global symmetries of the action (\ref{WZWint}). A first continuous symmetry of model (\ref{WZWint})  is PSU($N$) $=$ SU($N$)/ ${\mathbb{Z}}_N$ which acts as $G \rightarrow V G V^{\dagger}$, $V$ being an SU($N$) matrix. The center group of SU($N$), $V  \rightarrow \omega V$, has no effect on the action on $G$ so that
PSU($N$)  is the correct continuous symmetry group of Eq. (\ref{WZWint}). On top of this global symmetry, action (\ref{WZWint}) is also invariant under the ${\mathbb{Z}}_N$ symmetry (\ref{translation}) which corresponds to the one-step translation symmetry T$_{a_0}$ as well as under the ${\mathbb{Z}}_2$ charge conjugation $G \rightarrow G^{*}$. 

In the strong-coupling regime $g_n \rightarrow + \infty$, the potential term of Eq. (\ref{WZWint}) selects a SU($N$) matrix $G$ such that ${\rm Tr} \left[ G^n\right] = 0$ with $n=1,\ldots, \left[N/2\right]$. As shown in Ref. \onlinecite{seiberg}, the latter condition can be extended to $n=1,\ldots, N - 1$.  The eigenvalues of the $G$ matrix are, thus, proportional to the $N$th roots of unity and the fundamental WZNW SU($N$) $G$  field can be written as
\begin{eqnarray}
G &=& U \Omega U^{\dagger} \nonumber \\
\Omega &=&  \omega^{- (N-1)/2} 
\begin{pmatrix}
\omega^{N-1}  & 0 & \cdots & 0 \\
0 & \omega^{N-2}   & \cdots & 0 \\
 \vdots & \cdots & \omega & 0\\
0 & \cdots & 0 & 1
\end{pmatrix} , \label{wzwfieldmanifold}\\
\nonumber  
\end{eqnarray}
$U$ being a general U($N$) matrix. We, then, introduce $N^2$ complex scalar fields $\Phi_{i j}$ ($i,j = 1, \ldots, N$) 
such that $ U_{i j} = \Phi_{i j} = (\vec \Phi_j)_i$. These fields are constrained to be orthonormal complex vectors: $\vec \Phi^{*}_i \cdot \vec \Phi_j = \delta_{ij}$ to enforce the U($N$) property: $ U^{\dagger} U = I$. The identification (\ref{wzwfieldmanifold}) reads, thus, as follows in terms of the scalar fields: 
\begin{equation}
G_{i j} = \sum_a  \Phi^{*}_{j a}   \Omega_{aa}  \Phi_{i a} .
\label{Gfieldident}
\end{equation}
A U(1)$^N$ redundancy in the description (\ref{Gfieldident})  is manifest since the transformation $\vec \Phi_a \rightarrow e^{ i \theta_a} \vec \Phi_a$ gives the same $G_{i j}$ for all $\theta_a$ ($a =1, \ldots, N$). Distinct scalar fields take, thus, value in U($N$)/U(1)$^N$ $\sim$ SU($N$)/U(1)$^{N-1}$, i.e., the flag manifold \cite{bykov,bykov2,seiberg, tanizaki}.  
 
The original global symmetries of action (\ref{WZWint}) have a direct interpretation on the complex fields 
$\vec \Phi_i$ thanks to the identification (\ref{Gfieldident}). The PSU($N$) symmetry acts as  $ \Phi_{i j}   \rightarrow \sum_k V_{ik} \Phi_{k j}$, $V$ being an SU($N$) matrix. The one-step translation symmetry T$_{a_0}$ becomes $\vec \Phi_i \rightarrow \vec \Phi_{i+1}$ with $\vec \Phi_{N+1} = \vec \Phi_1$, whereas $\vec \Phi_i \rightarrow \vec \Phi^{*}_{N-i+1}$ corresponds to  the ${\mathbb{Z}}_2$ charge conjugation.

The next step of the approach is to replace the identification (\ref{Gfieldident}) in the action (\ref{WZW}) to derive the low-energy effective field theory for the complex fields $\vec \Phi_i$. The action, then, takes the form of a nonlinear $\sigma$ model on the flag manifold  SU($N$)/U(1)$^{N-1}$ with topological $\theta$ terms with a Lagrangian density \cite{seiberg, tanizaki},  
\begin{eqnarray}
&& {\cal L} = \frac{N}{4\pi} \sum_{a=1}^{N} \left( |\partial_{\mu}  {\vec \Phi_{a}} |^2
-  |{\vec \Phi^{*}}_{a} \cdot \partial_{\mu}  {\vec \Phi_{a}} |^2  \right)   \nonumber \\
&& \; \; \; \; \; \; +  \sum_{a=1}^{N}  \frac{\theta_a}{2\pi} \epsilon^{\mu \nu} 
\partial_{\mu}  {\vec \Phi^{*}}_{a} \cdot \partial_{\nu}  {\vec \Phi_{a}}
\label{flagsigma}  \\
&+&  \sum_{1 \le a < b \le N}  \left(g_{ab} \delta^{\mu \nu}  +  b_{ab} \epsilon^{\mu \nu} \right) \left({\vec \Phi^{*}}_{a} \cdot \partial_{\mu}  {\vec \Phi_{b}}  \right) \left( {\vec \Phi^{*}}_{b} \cdot \partial_{\nu}  {\vec \Phi_{a}} \right),
\nonumber 
\end{eqnarray}
with $\theta_a = 2\pi a$ ($a = 1, \ldots, N$), $g_{ab} = N \cos( 2\pi (a-b)/N)/2\pi$ and $b_{ab}  
= N \sin( 2\pi (a-b)/N)/2\pi$. Model (\ref{flagsigma}) contains $N$ topological angles $\theta_a$ with topological charges,
\begin{equation}
q_{a} = \frac{i}{2\pi}   \int d^2 x  \epsilon^{\mu \nu} 
\partial_{\mu}  {\vec \Phi^{*}}_{a} \cdot \partial_{\nu}  {\vec \Phi_{a}} 
\label{topocharges}
\end{equation}
which are integers. However, the topological charges are not all independent due to the orthonormalization constraint: 
$\vec \Phi^{*}_i \cdot \vec \Phi_j = \delta_{ij}$ and satisfy $\sum_{a=1}^{N} q_{a} = 0$ since it can be shown 
$\sum_{a=1}^{N} \vec \Phi^{*}_{a} \cdot \partial_{\mu}  {\vec \Phi_{a}}  = 0$\cite{tanizaki,affleckmila1}.  
It implies that model (\ref{flagsigma}) is left invariant by shifting all topological angles by a same amount: $\theta_a \rightarrow \theta_a + \theta$ for all $a$.
There are, thus, $N-1$ independent topological angles $\theta_a = 2\pi a$ ($a = 1, \ldots, N-1$) in model (\ref{flagsigma})
in full agreement with the value of the second homotopy group for the flag manifold $\Pi_2$ (SU($N$)/U(1)$^{N-1}) = \mathbb{Z}^{N-1} $. 

It has been shown recently that flag $\sigma$ model (\ref{flagsigma}) with topological angles  $\theta_a = 2\pi p a/N$ control the infrared properties of  SU($N$) Heisenberg spin chain (\ref{Heisenberg}) in symmetric rank-$p$ tensor representation in the large $p$ limit \cite{affleckmila,affleckmila1}. 
A gapless phase in the SU($N$)$_1$ universality class has been predicted for model (\ref{flagsigma}) when $p$ and $N$ are coprime whereas a spectral gap is formed in other situations \cite{seiberg,affleckwamer,tanizaki}. 
We, thus, expect that the perturbed CFT (\ref{WZWint}) is a massive field theory in the far-infrared regime since $p=N$ here. 

\section{Mapping onto Majorana fermions}

The deviation from the AJ integrable PSU($N$) spin model described by the Hamiltonian (\ref{modelcont}) corresponds to a fully gapped phase as seen from its relationship to the flag $\sigma$ model (\ref{flagsigma}) with $N-1$ independent topological angles $\theta_a = 2\pi a$ ($a = 1, \ldots, N-1$). In this section, we investigate directly the main physical properties of model (\ref{modelcont}) by exploiting a mapping onto Majorana fermions where we show explicitly its massive behavior.

\subsection{Conformal-embedding approach}

The infrared properties of model (\ref{modelcont}) strongly depend on the sign of the coupling constant $g$.
When $g<0$ , the minimization of the potential term $g {\rm Tr}  \Phi_{\rm adj} $ in Eq. (\ref{modelcont}) gives  $ G = e^{ 2 i k \pi/N} I = \omega^k I$ ($k=1, \ldots, N$) to maximize ${\rm Tr} G$ [see Eq. (\ref{tradj})]. The one-step translation ${\mathbb{Z}}_N$  symmetry (\ref{translation}) is, thus, spontaneously broken which signals the formation of a gapped phase with a $N$-fold degeneracy. When $g>0$, the nature of the ground state of model (\ref{modelcont}) is not as straightforward.  We show below that an 
SPT phase can show up in model (\ref{modelcont})  with $g>0$.

To this end, we exploit a conformal embedding which enables us to simplify model (\ref{modelcont}). The SU($N$) group is known to be a subgroup of Spin($N^2 -1$), the fundamental covering of the  SO($N^2 -1$) group. The central charge of the SU($N$)$_N$ CFT is $c= (N^2 - 1)/2$ which is that of the SO($N^2 -1$)$_1$ CFT  \cite{dms}.
This conformal embedding has been known from a long time \cite{olive} and has been fruitful, for instance, to investigate some 1D strange metals \cite{sachdev}.
The SO($N^2 -1$)$_1$ CFT spectrum admits several conformal towers defined by the integrable representations of its affine algebra: the identity,  vector, and spinor representations. If $N$ is odd, $N^2 -1$ is even, and there are two inequivalent spinor representations of dimension $2^{(N^2-3)/2}$: the spinor and its conjugate representation.
In contrast, when $N$ is even, $N^2 -1$ is odd and there is a single spinor representation of 
dimension $2^{(N^2-2)/2}$. The primary field transforming in the vector representation has $h_v = 1/2$ as conformal weight. In spinorial representations, this conformal weight  is $h_s = (N^2 -1)/16$.

The character decomposition of such conformal embedding for the Neveu-Schwartz sector
of the SO($N^2 -1$)$_1$ CFT is given by \cite{sachdev}
\begin{eqnarray}
\chi^{SO(N^2 -1)_1}_1 &=&  \chi^{SU(N)_N}_{(0 \dots 0)}  + \chi^{SU(N)_N}_{(20 \dots 10)} + 
 \chi^{SU(N)_N}_{(01 \dots 02)} + \ldots \nonumber \\
 \chi^{SO(N^2 -1)_1}_{\rm v} &=& \chi^{SU(N)_N}_{(10 \dots 01)} +  \chi^{SU(N)_N}_{(110 \dots 011)}  + \ldots,
\label{characterdecomp}
\end{eqnarray}
where $\chi^{SO(N^2 -1)_1}_{1,{\rm v}}$ are the SO($N^2 -1$)$_1$ character in the identity and the
vectorial representation of the SO($N^2 -1$) group, respectively. In Eq. (\ref{characterdecomp}), $\chi^{SU(N)_N}_{(10 \dots 01)}$ are SU($N$)$_N$  character in the SU($N$) representation labeled by their highest weights, here, $(10 \dots 01)$, for instance, (the adjoint representation). The SU($N$) representations with highest weights ${\Lambda} = (\lambda_1 \lambda_2 \ldots  \lambda_{N-1})$ appearing in the decomposition (\ref{characterdecomp}) satisfy the $N$-equality condition \cite{sachdev},
\begin{equation}
\sum_{i=1}^{N-1}  i  \lambda_i  = 0 \; \; {\rm mod}   \; N, 
\label{Nality}
\end{equation}
which means that the number of boxes $n_{Y}$ of the Young tableau of the SU($N$) representation is a multiple of $N$. Physically, it signals that the SU($N$)$_N$ fields that occur in the character decomposition (\ref{characterdecomp}) should be invariant under the one-step translation symmetry (\ref{translation}). In particular,  the fields involved in the perturbed CFT (\ref{modelcont}) and (\ref{WZWint}) are expressed only in terms of operators in the identity and
vector conformal towers of SO($N^2 -1$)$_1$ CFT.

The SU($N$)$_N$ adjoint primary field has scaling dimension $\Delta_{\rm adj} =1$. It corresponds to the SO($N^2 -1$)$_1$
primary field in the vectorial representation according to Eq. (\ref{characterdecomp}). In this respect, we introduce $N^2 - 1$ left-right-moving Majorana fermions $\xi^{A}_{R,L}$ normalized such that 
\begin{equation}
\xi^{A}_L (z) \xi^{B}_L (0)  \sim \frac{\delta^{AB}}{2 \pi z} .
\label{majoranaOPE}
\end{equation}
$A,B = 1, \ldots, N^2-1$ with a similar definition for the right-moving Majorana fermions. The adjoint SU($N$)$_N$ primary field (\ref{adjointfield}) has a simple free-field representation in terms of these fermions,
\begin{equation}
\Phi^{AB}_{\rm adj} \sim  - i \xi^A_R \xi^B_L .
\label{adjointfieldSUN}
\end{equation}
Model (\ref{modelcont}) can then be refermionized
\begin{equation}
{\cal H} = - \frac{iv}{2} \sum_{A=1}^{N^2-1} \left(\xi^{A}_{R} \partial_x  \xi^{A}_{R} - \xi^{A}_{L} \partial_x  \xi^{A}_{L} \right)
- i m\sum_{A=1}^{N^2-1} \xi^{A}_{R} \xi^{A}_{L}, 
\label{majoadjoint}
\end{equation} 
where we have neglected subleading marginal four-fermion contributions. 

Model (\ref{majoadjoint}) describes decoupled $N^2 - 1$ Majorana fermions with mass $m \sim g$. 
For all signs of $g$, the field theory is, thus, massive. For $m>0$, a nondegenerate fully gapped phase emerges.  This Majorana mapping constitutes the generalization of the Majorana approach of Ref. \onlinecite{tsvelik} to investigate the Haldane phase of the spin-1 Heisenberg chain starting from the SU(2)$_2$ critical point of the BT model \cite{babujian}.

\subsection{Edge states and SPT phases}

We now investigate the possible stabilization of a PSU($N$) SPT phase when $m >0$ by studying its edge excitations. To this end, model (\ref{majoadjoint}) is considered in a semi-infinite line with an open-boundary condition on $x=0$,
\begin{equation}
H = 
\frac{1}{2} \int_{0}^{\infty} dx \;\sum_{A=1}^{N^2-1}
\Psi^{A} \left(x\right)^{T} \left( -i v \sigma_3 \partial_x + 
m \sigma_2 \right) \Psi^{A} \left(x\right),
\label{hamtoy}
\end{equation}
where $\sigma_i$'s are the usual Pauli matrices and 
\begin{equation}
\Psi^{A} \left(x\right) =
\left(\begin{array}{c} \xi^{A}_{R} \left(x\right) \\ 
\xi^{A}_{L} \left(x\right) \end{array} \right).
\end{equation}
In our convention, the Majorana fermions are subject to the following boundary condition on $x=0$:
\begin{equation}
\xi^{A}_{R}  \left(0\right) = \xi^{A}_{L}  \left(0\right) ,
\label{BC}
\end{equation}
for all $A=1, \ldots, N^2-1$.
The Hamiltonian (\ref{hamtoy}) is exactly solvable
being quadratic in terms of the fermions and the 
resulting eigenvectors read as follows \cite{orignac}:
\begin{eqnarray}
\left(\begin{array}{c} \xi_R^{A}\left(x\right)  \\ \xi_L^{A}\left(x\right) \end{array} \right) &=& \frac 1 {\sqrt{2L}} \sum_{k>0} \left[
\xi^{A}_k \left(\begin{array}{c} \cos
\left(kx+\theta_k\right) + i \sin\left(kx\right) \\  \cos
\left(kx+\theta_k\right) - i \sin\left(kx\right) 
\end{array} \right)   \right. \nonumber \\
&+& \left. H.c.   \right]  + \sqrt{\frac{m}{v}} \left(\begin{array}{c} 1 \\ 1 \end{array}
\right) e^{-m x/v }\; \theta\left(m\right) \eta^A, 
\label{decomp}
\end{eqnarray}
where $\xi^{A}_k$ is a fermion annihilation operator
with wave-number $k=\pi n/L$, $ L$ being the large size of the line and $\theta$ is the Heaviside step function.
In Eq. (\ref{decomp}),  $\theta_k$ is given by
\begin{eqnarray}\label{eq:bogoliubov_rotation}
\cos \theta_k &=& \frac{vk}{\epsilon_k} \nonumber\\
\sin \theta_k &=& \frac{m}{\epsilon_k} ,
\end{eqnarray}
$\epsilon_k =\sqrt{v^2 k^2 +m^2}$ being the energy dispersion. The last term of Eq. (\ref{decomp}) is a zero-energy eigenvector of the Hamiltonian (\ref{hamtoy}) and, thus, the solution of both equations 
\begin{equation}
v \partial_x \xi^{A}_{R}  + m \xi^{A}_{L} = 0  \quad \mathrm{and} \quad
v \partial_x \xi^{A}_{L}  + m \xi^{A}_{R} = 0 .
\label{zeroenergycond}
\end{equation}
According to the boundary condition (\ref{BC}), this system gives a normalized solution if only if $m > 0$: $\xi^{A}_{R} (x)  = \xi^{A}_{L} (x) = \sqrt{m/v} \; e^{-m x/v }  \eta^A,$ with the normalization $\{\eta^A,  \eta^B \} =  \delta^{AB}$.  It signals the existence of $N^2-1$ exponentially Majorana localized states inside the gap (midgap states) for a positive mass $m$.

When $m>0$, $N^2 - 1$ Majorana zero-modes $\eta^{A}$, thus, emerge at the boundary of a semi-infinite chain and these edge states might give rise to some interesting 1D SPT phase. In the $N=2$ case, these three local Majorana modes form the generators $\Gamma^{AB} = i \eta^A \eta^B$ in the spinorial representation of the rotation group SO(3). They describe the spin-1/2 edge excitation of the Haldane phase \cite{orignac}.
For general $N$, not all these Majorana SPT phases, found in a continuum description, are actually protected by interactions. 
In particular, as recalled in the Introduction, it has been shown in Refs. \onlinecite{fidkowski,kitaev,turner} that 
time-reversal 1D Majorana topological phases are characterized by a ${\mathbb{Z}}_8$ classification in the  presence
of interactions. It means that time-reversal gapful phases with $k$-boundary Majorana modes modulo eight turns are equivalent \cite{kitaev,turner}.
When $N$ is odd, we have $N^2 - 1= 0 \; \; { \rm mod }   \; 8$, and
the topological phases, described by Eq. (\ref{majoadjoint}) with $m>0$ are, thus, not stable with interactions and adiabatically connected to a featureless nondegenerate gapful phase by adding four-fermion interactions.  
In contrast, model (\ref{majoadjoint}) with even $N$ have an odd number of robust Majorana zero modes and should describe a PSU($N$) SPT phase. 

The $N^2 - 1$ Majorana zero-modes $\eta^{A}$ also fix the projective representation of the SU($N$) group at the edge. This representation transcribes the physics of the PSU($N$) Heisenberg antiferromagnetic chain (\ref{Heisenberg}) in the symmetric rank-$N$ tensor representation. In this respect, let us introduce the following operator:
\begin{equation}
{\cal S}^{A} = - \frac{i}{2} f^{ABC} \eta^B \eta^C .
\label{majospinop}
\end{equation} 
It is straightforward to show that the operator (\ref{majospinop}) satisfies the 
SU($N$) algebra: $[{\cal S}^{A} , {\cal S}^{B} ] = i f^{ABC}  {\cal S}^{C}$.  The value of the corresponding quadratic Casimir operator suffices, here, to identify the SU($N$) irreducible representation of ${\cal S}^{A}$. A direct calculation gives
\begin{equation}
\sum_{A=1}^{N^2-1} {\cal S}^{A} {\cal S}^{A}  = \frac{N (N^2 -1)}{8},
\label{Casimir}
\end{equation} 
where we have used the following identities for the structure constants of the SU($N$) group:
\begin{eqnarray}
 f^{ABC} f^{ADE}  &=& \frac{2}{N}\left( \delta^{BD} \delta^{CE}  - \delta^{BE}  \delta^{CD}  \right) \nonumber \\
 &+& d^{ABD} d^{ACE} - d^{ABE} d^{ACD}, \nonumber \\
 d^{ABC} d^{ABC} &=& \frac{(N^2 -1)(N^2 -4)}{N}
\label{SUNident}
\end{eqnarray} 
$ d^{ABC}$ being the symmetric structure constants of the SU($N$) group. 
For $N=2$, Eq. (\ref{majospinop}) is the three-Majorana representation of a spin-1/2 operator
described in Ref. \onlinecite{tsvelikMajorana}. In the $N=3$ case, the operator  (\ref{majospinop}) corresponds to an SU(3) spin which belongs to the adjoint representation with quadratic Casimir $C_2({\rm adj}) = 3$. For $N=4$, the quadratic Casimir (\ref{Casimir}) is $15/2$ corresponding to the self-conjugate SU(4) representation $\mathbf{64}$ such that
\begin{equation}\label{eq:tableau64}
\mathbf{64} \longleftrightarrow \yng(3,2,1) .
\end{equation}

For general $N$, the edge state belongs to 
the SU($N$) representation with highest-weight $(1\ldots1)$ of dimension $2^{N(N-1)/2}$. The corresponding Young tableau has a number of boxes,
\begin{equation}
n_{Y\rm edge} = \sum_{i=1}^{N-1} i \lambda_i =  \sum_{i=1}^{N-1} i = \frac{N(N-1)}{2},
\label{edgestatPSUN}
\end{equation} 
and a quadratic Casimir given by Eq. (\ref{Casimir}).

Now, we can make contact with the cohomology classification of the PSU($N$) $\sim$ SU($N$)/${\mathbb{Z}}_N$ SPT phases \cite{quella}.
As recalled in the Introduction, there are $N-1$ topologically distinct  SPT phases. The inequivalent projective representations of PSU($N$) are labeled by a Young tableau where the number of boxes $n_{Y\rm edge}$ is defined modulo $N$.  The low-energy Majorana approach  (\ref{majoadjoint}) to PSU($N$) Heisenberg spin chain (\ref{Heisenberg}) in symmetric rank-$N$ tensor representation predicts a nondegenerate fully gapped phase with edge states characterized by $n_{Y\rm edge}  =N(N-1)/2$. When $N$ is odd, $n_{Y\rm edge}  = 0 \; {\rm mod}   \; N$ so that the phase is not a SPT phase. This observation agrees with our previous discussion related to the ${\mathbb{Z}}_8$ classification of interacting time-reversal phases of Majorana SPT phases. It also agrees for $N=3$ with the recent numerical investigation  of the Heisenberg model (\ref{Heisenberg}) in the three-box symmetric representation \cite{nataf2020}.
A nondegenerate phase with a very small gap $\Delta \simeq 0, 02 J$ has been reported whereas a critical behavior was more likely in previous numerical studies \cite{greiter,nataf2016}.  The edge states were also found to belong to the adjoint representation of SU(3) as expected from the underlying AKLT construction. \cite{rachelgreiter,rachelgreiter2,gozelmila} 
The Majorana field-theory approach for $N=3$, thus, reproduces these numerical and AKLT results.
Since $n_{Y\rm edge} = 3$, these edge states are not protected. The underlying phase is not an SPT phase in close parallel to the spin-2 Haldane phase which is not topologically protected \cite{oshikawapollmann}.
In contrast, when $N$ is even, the low-energy Majorana approach  (\ref{majoadjoint}) leads to SPT phases with topologically protected edge state which belong to the class: $n_{\rm top}  = N/2 \; {\rm mod}   \; N$  as seen from Eq. (\ref{edgestatPSUN}). The resulting PSU($N$) SPT phase belongs, thus, to the same topological class as the SPT phase of the 
Heisenberg spin chain where the spin transforms in the self-conjugate representation (\ref{sptnonne}). The latter SPT phase has edge states transforming in the self-conjugate antisymmetric representation  \cite{Nonne2013,bois,totsuka},
\begin{equation*}
\text{\scriptsize $N/2$} \left\{ 
\yng(1,1,1,1)
 \right.  .
\end{equation*}
In the simplest $N=4$ case, the Appendix provides the AKLT model for the Heisenberg spin chain (\ref{Heisenberg}) in the four-box  fully symmetric representation inspired by the AKLT construction of Ref. \onlinecite{gozelmila}. The edge states of the AKLT SPT phase are shown to belong to the SU(4) representation with the Young tableau (\ref{eq:tableau64}) with a dimension of 64 like in the Majorana-fermion approach. For general $N$, the same construction also exists for spins in the $N$-box symmetric representation with edge states belonging to the SU($N$) representation with highest-weight (1,1,...1). The phase of the model obtained is also expected to be topological only in the $N$ even case, such as in the Majorana approach. 

The  nonlinear $\sigma$ model (\ref{flagsigma}) on the flag manifold  SU($N$)/U(1)$^{N-1}$ with topological terms $\theta_a = 2\pi a$ ($a = 1, \ldots, N-1$) in an open geometry should also reveal the nature of the edge state of the underlying SPT phase. In the simplest case $N=2$, the $\sigma$ model is the CP$^{1}$ model with a $\theta = 2 \pi$ term and it has been shown that spin-1/2 edge states emerge in an open geometry as it should be to describe the Haldane phase of the spin-1 Heisenberg chain \cite{ng}. For general $N$,  we expect that the $\sigma$ model on the flag manifold with $\theta_a = 2\pi a$ has edge states whose representation under SU($N$) is encoded by the values of its topological angles. We conjecture that the length of the $a$th row of the Young tableau of the representation of the edge state is $l_a = \theta_{N-a}/2 \pi= N-a$ with $a = 1, \ldots, N-1$ so that the highest weight of the SU($N$) representation is $\Lambda_{\rm edge} = (1\ldots1)$ as found within the Majorana approach.

\section{Concluding remarks}

In this paper, we have presented a Majorana fermion approach to investigate the possible formation of a PSU($N$) SPT phase
as the ground state of the  Heisenberg spin chain (\ref{Heisenberg}) in the $N$-box  fully symmetric representation. By exploiting the existence of the AJ integrable spin model with SU($N$)$_N$ quantum critical behavior, we describe the fully gapped phase of model (\ref{Heisenberg}) by means of $N^2-1$ noninteracting massive Majorana fermions. This approach is the generalization of the one for $N=2$, proposed in Ref. \onlinecite{tsvelik}, which accounts for the Haldane phase of the spin-1 Heisenberg chain in terms of three massive Majorana fermions.  

Our paper enables the determination of the underlying SPT phase of model (\ref{Heisenberg}) through its edge states encoded in $N^2-1$ Majorana zero modes. When $N$ is odd, we find that the nondegenerate phase gapful phase is not topologically protected and, thus, equivalent to a featureless phase. In contrast, the SPT phase with even $N$ is protected by the PSU($N$)  symmetry and belongs to the same topological class as the  PSU($2n$)  SPT phase with the edge state in the self-conjugate fully antisymmetric representation. After the spin-1 Haldane phase, the simplest SPT phase of the Heisenberg spin chain (\ref{Heisenberg}) in the fully symmetric representation is obtained for $N=4$.  Both the edge states of this phase and the edge states in the ${\bf 6}$ representation of the SU(4) group belong to the same topological class. A numerical investigation, using similar tools as in Refs. \onlinecite{nataf2018,nataf2020,ladderphle}, is naturally called for to confirm this prediction, obtained within a low-energy description. The SPT phase for N=4 can also be explored in a four-leg  SU(4) spin ladder where the spins belong to the fundamental representation of SU(4). For a sufficiently strong ferromagnetic interchain coupling, the SPT phase is expected to emerge. This ladder system can be realized by considering, for instance, ytterbium atoms in their ground state by keeping only four nuclear states to realize the SU(4) symmetry  \cite{cazalillaho,rey,cazalilla,taie,fallani,ye,folling}.
The four-ladder geometry can be obtained, in principle, by selective evaporation of anisotropic collection of 1D tubes made of ytterbium atoms. However, as for the Haldane phase,  the temperature scale to reach the SPT physics will be difficult to reach in actual cold atom experiments as well as the realization of a ferromagnetic interchain coupling between the tubes.

As a perspective, it will be interesting to further generalize our CFT approach to investigate the degenerate gapped phases of model (\ref{Heisenberg}) for representations with a number of box $n_Y= pN$. We hope to come back to this issue elsewhere.

\begin{acknowledgements}
The authors are very grateful to S. Capponi and K. Totsuka for important discussions and collaboration over the years on this topic. They also acknowledge  D. Schuricht, T. Dupic, and A. A. Nersesyan for useful discussions. P.F. was supported by the European Research Council under Grant No. 758329 (AGEnTh), and has received funding from the European Union's Horizon 2020 Research and Innovation Programme under Grant Agreement No 817482 (PASQuanS).
\end{acknowledgements}


\appendix
\section{Appendix: AKLT construction for the $N=4$-box fully symmetric representation}

In this Appendix, we present an AKLT construction of the Heisenberg spin chain (\ref{Heisenberg}) in the four-box  fully symmetric representation following the method presented in Ref.~\onlinecite{gozelmila}. This construction confirms the identification of the edge state found in the Majorana fermion approach.

The construction of Ref.~\onlinecite{gozelmila} generalizes the construction of the AKLT chain for $N=2$. Let $\mathcal{P}$ be the representation of the on-site physical SU($N$) spin, and $\mathcal{V}$ be a self-conjugate representation of a virtual SU($N$) spin. The construction consists in dividing $\ell$ physical spin transforming in $\mathcal{P}^{\otimes\ell}$ into two virtual spins from $\mathcal{V}$ (and $\ell-1$ singlets) projected into $\mathcal{P}^{\otimes\ell}$. The resulting chain is gapped with a unique ground state for periodic boundary conditions, and with degenerate ground states for open-boundary conditions. In both cases, the bulk of each state displays the same nonmagnetic coupling of $\ell$ neighboring spins into singlets. The degeneracy of the open-boundary chain comes from the edge states with each edge transforming in the representation $\mathcal{V}$.

The construction of Ref.~\onlinecite{gozelmila} is possible under two conditions. First, the decomposition into irreducible representation of the tensorial product of the self-conjugate $\mathcal{V}$ with itself must contain the physical irreducible representation $\mathcal{P}$, possibly with multiplicity, i.e., $\mathcal{P} \in \mathcal{V} \otimes \mathcal{V}$. Second, two virtual spins must be able to form a singlet. This condition is always verified when $\mathcal{V}$ is self-conjugate. Therefore, this construction is possible for $N=4$ with
\begin{equation}
    \mathcal{P}= \yng(4) \;\; \text{and} \;\; \mathcal{V}= \yng(3,2,1). \label{eq:defPV}
\end{equation}
Indeed, $\mathcal{P}$ appears in the decomposition of $\mathcal{V} \otimes \mathcal{V}$,
\begin{equation}
    \begin{split}
        \yng(3,2,1) \otimes \yng(3,2,1) = \bullet  \oplus 3\times\yng(2,1,1) \oplus 2\times\yng(2,2) \\
        \oplus \yng(4) \oplus \yng (4,4,4) \\
        \oplus 3\times\yng(3,1) \oplus 3\times\yng(3,3,2) \oplus 3\times\yng(4,2,2) \\
        \oplus \yng(4,4) \oplus 4\times\yng(4,3,1) \\
        \oplus 2\times\yng(5,4,3) \oplus 2\times\yng(5,2,1) \\
        \oplus \yng(5,3) \oplus \yng(5,5,2) \\
        \oplus \yng(6,3,3) \oplus \yng(6,4,2).
    \end{split}\label{eq:VxV}
\end{equation}
Using the Young tableaux' respective dimensions, the decomposition reads:
\begin{equation}
\begin{split}
\mathbf{64}\otimes \mathbf{64}=& \mathbf{1} \oplus 3\times\mathbf{15} \oplus 2\times\mathbf{20} \oplus \mathbf{35} \oplus \mathbf{\overline{35}} \\
& \oplus 3\times\mathbf{45} \oplus 3\times\mathbf{\overline{45}} \oplus 3\times\mathbf{84} \oplus \mathbf{105} \\
& \oplus 4\times\mathbf{175} \oplus 2\times\mathbf{256} \oplus 2\times\mathbf{\overline{256}} \\
&\oplus \mathbf{280} \oplus \mathbf{\overline{280}} \oplus \mathbf{300} \oplus \mathbf{729}.
\end{split}
\end{equation}
The decomposition of $\mathcal{V} \otimes \mathcal{V}$ follows the standard rules of the tensor product~\cite{ITZYKSON1966}.
The dimension $D_N$ of the SU($N$) representation of Young tableau $Y$ with rows of $\lbrace l_1, l_2, \dots, l_{N-1}, l_N=0 \rbrace$ boxes (in descending order) reads as follows~\cite{ITZYKSON1966}: 
\begin{equation}
    D_N\left(Y \right)= \frac{\prod_{1\leq i < j \leq N} \left(l_i-l_j+j-i \right)}{\left(N-1\right)!\left(N-2\right)!\dots 1!}.
\end{equation}

In this case, there exists a parent Hamiltonian coupling only nearest neighbor ($\ell=2$) whose ground states are edge states transforming in $\mathcal{V}$. Indeed, the decomposition of $\mathcal{P}\otimes \mathcal{P}$ is as follows:
\begin{equation}
\multlinegap=3cm
    \begin{multlined}
        \yng(4) \otimes \yng(4) = \\
        \yng(4,4) \oplus \yng(8) \\
        \oplus  \yng(5,3) \oplus \yng(7,1) \\
         \oplus \yng(6,2).
    \end{multlined}\label{eq:PxP}
\end{equation}
Using the dimension, the decomposition reads
\begin{equation}
    \mathbf{35} \otimes \mathbf{35} = \mathbf{105} \oplus \mathbf{165} \oplus \mathbf{280} \oplus \mathbf{315} \oplus \mathbf{360}.
\end{equation}
The representation $\mathbf{105}$ and $\mathbf{280}$ are found in both Eqs.~(\ref{eq:VxV}) and (\ref{eq:PxP}). Thus, the Hamiltonian on two physical spin in $\mathcal{P}^{\otimes 2}$ that equally favors the two representations $\mathbf{105}$ and 
$\mathbf{280}$, is
\begin{equation}
    h= I -\mathbb{P}_{\mathbf{105}}-\mathbb{P}_{\mathbf{280}}, 
 \label{eq:2spinhMPS}
\end{equation}
where $\mathbb{P}_{\mathbf{105}}$ and $\mathbb{P}_{\mathbf{280}}$ are the projectors onto the representation $\mathbf{105}$ and $\mathbf{280}$ respectively.
There are two ways to interpret the 105+280=385 ground states of this two-sites Hamiltonian. The first way sees the two physical spins of $\mathbf{35}$ align such that only the superpositions transforming in $\mathbf{105} \oplus \mathbf{280}$ are ground states of the system. The second way divides each physical spin into two virtual ones from $\mathbf{64}$ projected back into  $\mathbf{35}$. The Hamiltonian (\ref{eq:2spinhMPS}) favors energetically the coupling of two neighboring virtual spins of different sites into a singlet, leaving free the two virtual spins on the edge of the system. Because of the initial projection of the Hilbert space on each site into $\mathbf{35}$, the Hilbert space of the two free edge states is restricted to $\mathbf{105} \oplus \mathbf{280}$ only, instead, of all the representations in Eq.~(\ref{eq:VxV}). The AKLT-inspired parent Hamiltonian of the full open chain reads
\begin{equation}
    H_{\rm{AKLT}}= \sum_{i=1}^{L-1} \tau_i (h), \label{eq:chainhMPS}
\end{equation}
with $\tau_i$ the translation operator on site $i$ and $L$ the number of sites. The 385 ground states of this system can be interpreted as a chain of $2L$ virtual spins in $\mathbf{64}$. Two virtual spins of each neighboring physical site pair up into a singlet, such that only two virtual spins are left unpaired, one on each edge. Because of Hilbert space restrictions, the two edge spins together transform in $\mathbf{105} \oplus \mathbf{280}$ only. The latter restriction is incompatible with a semi-infinite chain such that 64 states can be expected on the one edge of this geometry.

The quadratic Casimir is enough to obtain an explicit expression for $\mathbb{P}_{\mathbf{105}}$ and $\mathbb{P}_{\mathbf{280}}$ and, hence, $H_{\text{AKLT}}$. The quadratic Casimir of a representation $\mathcal{R}$ of highest weight $\Lambda$ is
\begin{equation}
    C_2 (\mathcal{R})=\frac{1}{2}\langle \Lambda , \Lambda + 2 \sum_{i} \Lambda_i\rangle ,
\end{equation}
where $\Lambda_i$ are the fundamental weights. When $\mathcal{R}$ is an irreducible representation of SU($N$), $\Lambda=\sum_{i=1}^{N-1} \lambda_i \Lambda_i$ where $\lambda_i$'s are the Dynkin labels of the representation. In this case,
\begin{equation}
    \langle \Lambda_i, \Lambda_j \rangle = \text{min}(i,j) - \frac{ij}{N}.
\end{equation}
For the representation in the decomposition Eq.~(\ref{eq:PxP}), we find
\begingroup
\allowdisplaybreaks
\begin{align*}
    C_2(\mathbf{105})&= 16, \\
    C_2(\mathbf{165})&= 36, \\
    C_2(\mathbf{280})&= 18, \\
    C_2(\mathbf{315})&= 279/8, \\
    C_2(\mathbf{360})&= 22 . 
\end{align*}
\endgroup
We call $\mathbf{S_T}$ the total spin of the system of the two physical spins. If $\mathcal{A}_i$ are the irreducible representation in Eq.~(\ref{eq:PxP}) of  $\mathcal{P}^{\otimes 2}$ such that $\mathcal{P}^{\otimes 2}= \oplus_{i=1}^5\mathcal{A}_i $, the expressions of the projectors follow:
\begin{equation}
    \begin{split}
       \mathbb{P}_{\mathcal{A}_i}= \frac{1}{C_2(\mathcal{A}_i)}\prod_{j\neq i}\left( \mathbf{S_T}^2-C_2(\mathcal{A}_j)\right).
    \end{split}
\end{equation}
The parent Hamiltonian (\ref{eq:chainhMPS}), thus, involves a polynomial of degree 8 of the nearest-neighbors spin coupling.


\end{document}